\documentstyle[preprint,tighten,floats,aps,epsf]{revtex}

\begin{document}
\title
{Variation after Angular Momentum Projection for the Study of Excited 
States Based on Antisymmetrized Molecular Dynamics
}

\author{Y. Kanada-En'yo}

\address{Institute of Particle and Nuclear Studies, \\
High Energy Accelerator Research Organization,\\
3-2-1 Midori-cho Tanashi Tokyo 188, Japan}

\maketitle
\begin{abstract}
In order to study the structure of excited states
we perform a variational calculation after spin parity projection (VAP) 
within the framework of Antisymmetrized Molecular Dynamics (AMD).
The framework is proven to be a new powerful approach for 
the study of the various structures of excited states because it is free from 
model assumptions such as inert cores, existence of clusters,
and the axial symmetry.
By using finite range interactions 
with a density dependent term we reproduce well all the energy
levels below 15 MeV in $^{12}$C.
This is the first theoretical model that reproduces many 
$E2$ transition rates and 
$\beta$ decays to $^{12}$C successfully.

\end{abstract}


Clustering is one of the important features
of the light nuclei.
The Ikeda diagram \cite{IKEDAa} predicts molecule-like 
structures in the excited states, 
according to the activation of the inter-cluster relative motion.
The cluster model has been successful 
for the study of the molecule-like states.
However the model is not suitable for describing the 
shell-model-like features,
such as the $j$-$j$ coupling and the particle-hole excitations.
The model, as we see later, failed to reproduce 
important data such as the 
level spacing and the transition strength because 
the dissociation of clusters is not taken into account by
the conventional cluster model.
These shell-model-like aspects are often found 
in many states including the ground state.
On the other hand, the shell model successfully reproduces properties of
many excited levels of light nuclei (for example \cite{SHELL1,SHELL2}),
 but a number of states in light nuclei have been left unsolved because
it is difficult for the shell model
to describe well-developed molecule-like states which may exist
in low energy region.

There are very few approaches which can describe both
the molecule-like and
the shell-model-like phenomena.
For instance, in $^{12}$C which has been studied for a long time 
\cite{RGM,OCM,GCM,ITAGAKI}, 
the structure of the excited $0^+_2$, $0^+_3$, and $2^+_2$ 
 states is not yet clarified.
Therefore, it is desirable to develop a theoretical approach
that is able to deal with both aspects in a unified manner.
Our aim is to systematically study how the structure of nuclei changes
with increasing excitation energy. 

The AMD has already proved to be a very useful theoretical 
approach for the structure of general light nuclei 
\cite{ENYOa,ENYOb,ENYOc,ENYOd}.
The AMD basis wave functions are written as Slater determinants 
where the spatial part of each single particle wave function
is a Gaussian.
An important point of AMD is that we do not need model assumptions 
such as inert cores and 
the axial symmetry.
In the previous works on the ground states of light unstable nuclei
\cite{ENYOb,ENYOc,ENYOd}, 
we applied a simplest version of AMD in
which we make energy variation only after parity projection of
a Slater determinant.
The simplest version of AMD was found to describe
the properties of ground states successfully.
We also studied the change in structure along the yrast line of $^{20}$Ne 
by a cranking method of AMD \cite{ENYOa}
instead of the angular momentum projection before energy variation.

In this letter,
we propose a new microscopic approach of 
variational calculation after parity and total angular momentum
projection within the framework of antisymmetrized molecular dynamics(AMD).
In order to confirm the userfulness of the method,
we apply this approach for the first time
to the study of excited states of $^{12}$C
using finite range interactions.


Concerning the formulation of AMD, 
the reader is refered to papers
\cite{ENYOa,ENYOb,ENYOc}. An AMD wave function 
is a Slater determinant of Gaussian wave packets;

\begin{eqnarray}
&\Phi_{AMD}={1 \over \sqrt{A!}}
{\cal A}\{\varphi_1,\varphi_2,\cdots,\varphi_A\},\\
&\varphi_i=\phi_{{{\bf Z}}_i}\chi_i\tau_i :\left\lbrace
\begin{array}{l}
\phi_{{{\bf Z}}_i}({\bf r}_j) \propto
\exp\left 
[-\nu\biggl({\bf r}_j-{{\bf Z}_i \over \sqrt \nu}\biggr)^2\right],
\label{eqn:single}\\
\chi_{i}=
\left(\begin{array}{l}
{1\over 2}+Z_{4i}\\
{1\over 2}-Z_{4i}
\end{array}\right).
\end{array}\right.
\end{eqnarray}
where the centers of Gaussians ${\bf Z}_i$'s are complex variational
parameters. $\chi_i$ is the intrinsic spin function parametrized by
$Z_{4i}$ and $\tau_i$ is the isospin
function which is fixed to be up(proton) or down(neutron)
 in the present calculation.
We vary the energy expectation values for the 
parity and total angular momentum eigenstates (VAP calculation), 
\begin{equation}
{\langle P^J_{MK'}\Phi^\pm_{AMD}|H|P^J_{MK'}\Phi^\pm_{AMD}\rangle \over
\langle P^J_{MK'}\Phi^\pm_{AMD} |P^J_{MK'}\Phi^\pm_{AMD}\rangle },
\end{equation}
where the operator of total angular momentum projection $P^J_{MK'}$ is 
$\int d\Omega D^{J*}_{MK'}(\Omega)R(\Omega)$.
The expectation values are calculated numerically by the 
sum of mesh points instead of integral on Euler angle $\Omega$.
We make energy variation by the use of the frictional cooling method 
\cite{ENYOb}. 
Here we represent the intrinsic wave function $\Phi_{AMD}$ obtained by VAP
for the lowest states with a given spin parity $J^\pm$ as
$\Phi^{J\pm}_0$.
Higher excited states are constructed by superposing wave functions 
so as to orthogonalize to the lower states. That is to say that the 
$n$-th $J^\pm$ state $\Phi^{J\pm}_n$ is calculated by 
varing the energy expectation value of the
othogonal component to the lower states;
\begin{equation}
P^{J\pm}_{MK'}\Phi^{J\pm}_n-\sum^{n-1}_{k=1}
\langle P^{J\pm}_{MK'}\Phi^{J\pm}_k|P^{J\pm}_{MK'}\Phi^{J\pm}_n\rangle
P^{J\pm}_{MK'}\Phi^{J\pm}_k.
\end{equation}
By making VAP calculations with various sets of $\{J\pm,n\}$
 we obtain many intrinsic states $\{\Phi_1,\cdots,\Phi_m\}$, 
which approximately correspond to the $J^\pm_n$ states.
Final results are constructed by diagonalizing 
the hamiltonian matrix
$\langle P^{J\pm}_{MK'}\Phi_i|H|
P^{J\pm}_{MK''}\Phi_j\rangle$ formed from all the intrinsic states.
The resonance states are treated within a bound state approximation.

In this work the adopted interactions are the central force of 
the modified Volkov no1 of case 3 \cite{TOHSAKI} which consists of 
the finite range two-body force and 
the zero range three-body repulsive force as a density dependent term, 
the spin-orbit force of G3RS with two range Gaussians
\cite{LS} and the Coulomb force.
The Majorana parameter used here is $m=0.62$, 
and the strength of G3RS force is 
$u_1=-u_2=3000$ MeV. We chose the width parameter $\nu$ for Gaussians
in Eq.\ref{eqn:single} as 0.18 . 

\begin{figure}[htb]
\noindent
\epsfxsize=1.0\textwidth
\centerline{\epsffile{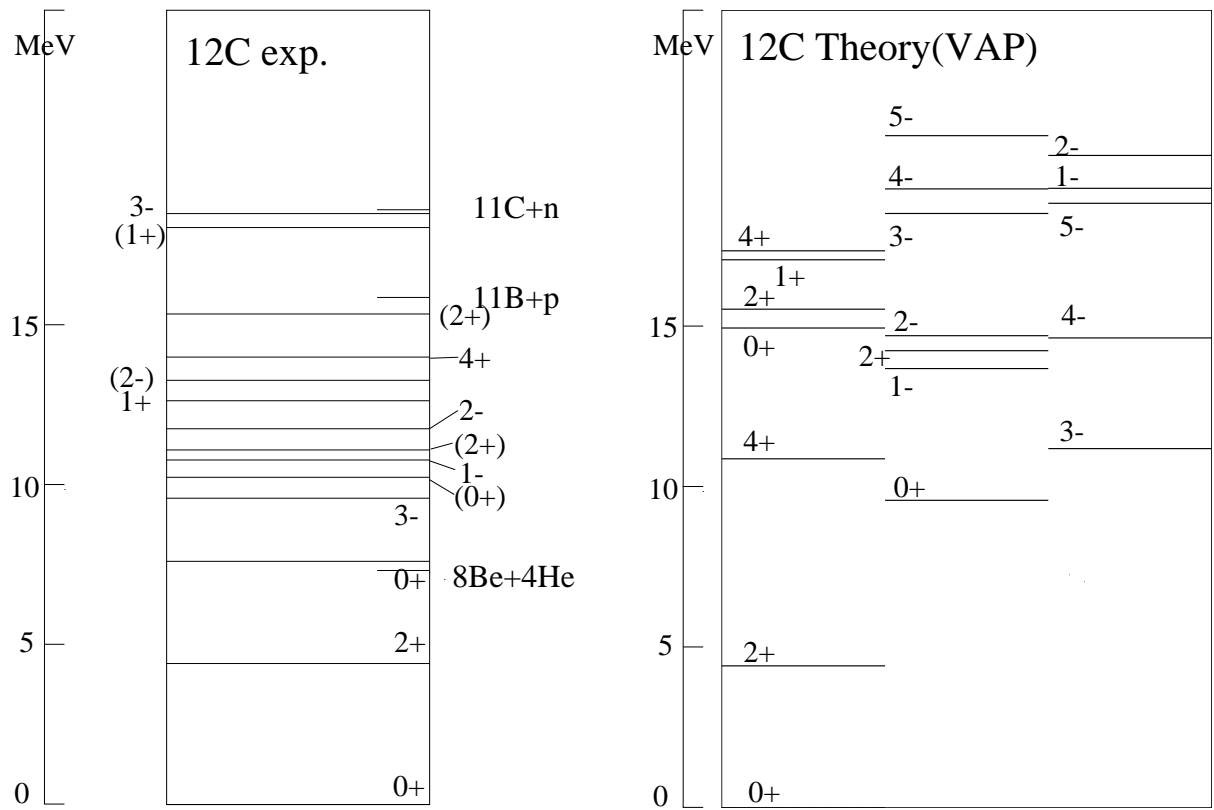}}
\bigskip

\caption{\label{fig:c12sped}
Excitation energies of the levels of $^{12}$C.
In the right-hand side, VAP calculations in the AMD framework 
are shown and compared with the experimental data (left).
}

\end{figure}

Energy levels of $^{12}$C are shown in Fig.\ref{fig:c12sped}
compared with the experimental data.
The AMD calculations reproduce
the energy levels very well. In the cluster model calculations  
the level spacing between $0_1^+$ and $2^+_1$ was always underestimated.
For instance, in a GCM calculation \cite{GCM} it was 2.2 MeV which is
much smaller than the experimental value 4.4MeV.  
In the present result the large level spacing between $0_1^+$ and
$2^+_1$ agree with the experimental data. We think that it is 
because the theory describes successfully
the dissociation of $\alpha$ cluster in the ground $0_1^+$ state 
due to the LS force. 
It is found that molecule-like states with well-developed 3$\alpha$
structures construct rotational bands of $K^\pi$=$3^-, 1^-, 0^+_2, 0^+_3$. 
Even by a $(0+2)\hbar\omega$ shell-model calculation for $p-shell$ nuclei
\cite{SHELL2} they can not describe $0^+_3$ and 
$2^+_2$ states in the $K^\pi=0^+_3$ band  because such states 
contain highly excited $n$ $\hbar \omega$ components in terms of 
the shell model.

The transition strengths are of great help 
to investigate the structure of excited states.
The theoretical and the experimental values of
$E2$ transition strengths are shown in Table \ref{tbl:c12e2}.
In a GCM calculation\cite{GCM} of the $3\alpha$ cluster model 
the $B(E2;0_2^+\rightarrow 2^+_1)$ was about one third of  
the experimental data (see Table \ref{tbl:c12e2}. The present result for 
$B(E2;0_2^+\rightarrow 2^+_1)$ is as much as the experimental data.

The calculated strengths of Gamov-Teller $\beta$ decay transitions from 
$^{12}$N to the excited states of $^{12}$C are shown 
in Table \ref{tbl:c12beta} and compared with the experimental data.
The ground state of the 
parent nucleus $^{12}$N is obtained with a variational calculation 
after spin parity projection to $1^+$ states.
It is difficult to discuss $\beta$ transitions to $^{12}$C$^*$ 
within the framework of the $3\alpha$ cluster model, because $\beta$ decay of 
nucleons in $\alpha$ clusters is forbidden. 
In other words, the origin of the $\beta$ decay strength is 
the components of the dissociation of the $\alpha$ clusters.
On the other hands, even in shell-model calculations with the large model 
space of $(0+2)\hbar\omega$ the $0^+_3$ state does not appear 
in low energy region as mentioned above. 
Therefore there has been no theoretical approach 
which reproduce the strength of 
$\beta$ decay to the excited $0^+$ states. 
This is the first microscopic calculation that can describe 
a large number of 
the experimental data of the transition strength to the excited states of
$^{12}$C.   
It means that in the present calculations the component of the dissociation 
of $\alpha$ clusters 
is reasonably large enough to explain the $\beta$ transition strength.
This success is due to the flexibility of AMD wave functions 
which can describe both shell-model-like and molecule-like aspects.

Another shell-model-like aspect is found in the 
 $1^+_1(T=0)$ state which is considered to be a shell-model-like state and 
can not be described in 3$\alpha$ cluster models.
In this work the $\beta$ transition strength to $1^+_1$ from $^{12}$N 
reasonably agrees to the experimental data for the $1^+(T=0)$ state 
at 12.71 MeV, though the excitation energy is overestimated
(see Table \ref{tbl:c12beta} and Fig \ref{fig:c12sped}).


In summary, we achieved the variation after the parity and 
the total angular momentum projection within the framework of 
antisymmetrized molecular dynamics and 
applied it to the excited states of $^{12}$C.
The framework is a new approach for the study of the structure 
of excited states
with freedom from any model assumptions such as 
inert cores and the axial symmetry.
It is a practical method of the variation after spin parity projection
with finite range interactions.
The model can represent both clustering structures 
and the dissociation of clusters due to shell-model-like aspects
within one framework. 
This microscopic model is a powerful tool which is 
applicable to stable as well as unstable nuclei.
Many excited levels 
of $^{12}$C, $E2$ transition strengths and 
$\beta$ decays to the excited states are reproduced within one microscopic 
framework for the first time. We can obtain information of the structures of 
many excited states from the experimental data of the transition strength.
It owes to the flexibility of the AMD wave function 
which can describe both shell-model-like and clustering aspects
seen in the excited states. 

The author would like to thank H. Horiuchi for many discussions and
advises. She is also thankful to N. Itagaki, Y. Kondo and
O. Morimatsu for helpful discussions. She is grateful to A. Ono
for technical advices on numerical calculations.
This work was partly supported by 
Research Center for
Nuclear Physics (RCNP) of Osaka University and Institute of Physical and
Chemical Research (RIKEN) for using computer facilities.

\begin{table}
\caption {\label{tbl:c12e2} The strength of the $E2$ transitions of
$^{12}$C.}
\begin{center}
\begin{tabular}{c|c|c|c}
Transition & exp. & present & GCM\cite{GCM} \\
\hline
$B(E2;2^+_1 \rightarrow 0^+_1)$ & 7.81$\pm$0.44 e$^2$fm$^4$ & 8.8 e$^2$fm$^4$
& 8.0 e$^2$fm$^4$ \\
$B(E2;0^+_2 \rightarrow 2^+_1)$ & 13.4$\pm$1.8 e$^2$fm$^4$ & 19.3 e$^2$fm$^4$
& 3.5 e$^2$fm$^4$ \\
\end{tabular}
\end{center}

\end{table}
\begin{table}
\caption {\label{tbl:c12beta} The experimental data for $\beta$ decays 
$^{12}$N($\beta^+$)$^{12}$C compared with the theoretical results.
}
\begin{center}

\begin{tabular}{cccc}
Decay to $^{12}$C$^*$ (MeV) & $J^\pi$ & \multicolumn{2}{c}{log$ft$} \\
& & exp.& theory ($J_f^\pm$) \\
\hline
0 & $0^+$ & 4.120 $\pm$ 0.003 & 3.8 ($0^+_1$) \\
4.44 & $2^+$ & 5.149 $\pm$ 0.007 & 4.8 ($2^+_1$)\\
7.65 & $0^+$ & 4.34 $\pm$ 0.06 & 4.0 ($0^+_2$)\\
10.3 & $(0^+)$ & 4.36 $\pm$ 0.17 & 4.7 ($0^+_3$)\\
12.71 & $1^+$ & 3.52 $\pm$ 0.14 & 3.8 ($1^+_1$)\\
\end{tabular}
\end{center}

\end{table}

\begin{thebibliography}{9}
\bibitem{IKEDAa}
K. Ikeda, N. Takigawa and H. Horiuchi, Prog. Theor. Phys. Supple.
 Extra Number, 464 (1968).
\bibitem{SHELL1}
A.G.M. van Hees and P.W.M. Glaudemans, Z. Phys. A {\bf 315}, 223 (1984). 
\bibitem{SHELL2}
A.A. Wolters, A.G.M. van Hees, and P.W.M. Glaudemans, 
Phys. Rev. C {\bf 42}, 2053 (1990).
\bibitem{RGM}
Y. Fukushima and M. Kamimura, 
{\it Proc. Int. Conf. on Nuclear Structure, Tokyo, 1977} 225
\bibitem{OCM}
H. Horiuchi, Prog. Theor. Phys. {\bf 51} (1974), 1266; {\bf 53}, 447 (1975)
\bibitem{GCM}
E. Uegaki, S. Okabe, Y. Abe and H. Tanaka, Prog. Theor. Phys. {\bf 57}, 
1262 (1977).
E. Uegaki, Y. Abe, S. Okabe and H. Tanaka, Prog. Theor. Phys. {\bf 59},
 1031 (1978); {\bf 62}, 1621 (1979).
\bibitem{ITAGAKI}
N. Itagaki, A. Ohnishi and K. Kat\=o, Prog. Theor, Phys. {\bf 94}, 1019
(1995).
\bibitem{ENYOa}
 Y. Kanada-En'yo, and H. Horiuchi, Prog. Theor. Phys.
{\bf 93} 115 (1995).
\bibitem{ENYOb}
 Y. Kanada-En'yo, H. Horiuchi, and A. Ono, 
Phys. Rev. C {\bf 52}, 628 (1995).
\bibitem{ENYOc}
 Y. Kanada-En'yo and H. Horiuchi,
Phys. Rev. C {\bf 52}, 647 (1995).
\bibitem{ENYOd}
 Y. Kanada-En'yo and H. Horiuchi,
 Phys. Rev. C {\bf 54}, R468 (1996).
\bibitem{TOHSAKI}
 T. Ando, K.Ikeda, and A. Tohsaki, Prog. Theor. Phys.
 {\bf 64}, 1608 (1980).
\bibitem{LS}
 N. Yamaguchi, T. Kasahara, S. Nagata, and Y. Akaishi,
 Prog. Theor. Phys. {\bf 62}, 1018 (1979);
 R. Tamagaki, Prog. Theor. Phys. {\bf 39}, 91 (1968).
\end{thebibliography}
\end{document}